\documentclass[conference]{IEEEtran}

\usepackage{graphicx}
\usepackage[center]{caption}
\usepackage{multirow}
\usepackage{textcomp}
\usepackage{adjustbox,lipsum}
\usepackage{xspace}
\usepackage{amsmath}
\usepackage{amssymb}
\usepackage{amsfonts}
\usepackage{amstext}
\usepackage{amsthm}
\usepackage{tabularx}
\usepackage{caption}
\usepackage{subcaption}
\usepackage{siunitx}
\usepackage{xcolor}
\usepackage{colortbl}
\usepackage[english]{babel}
\usepackage[utf8]{inputenc}
\usepackage{algorithm}
\usepackage[noend]{algpseudocode}
\usepackage{arevmath}     

\usepackage{subfiles} 

\newcommand{\MethodName}{CAN-LOC\xspace}

\algnewcommand\algorithmicforeach{\textbf{for each}}
\algdef{S}[FOR]{ForEach}[1]{\algorithmicforeach\ #1\ \algorithmicdo}

\begin{document}


\title{\MethodName : Spoofing Detection and Physical Intrusion Localization on an In-Vehicle CAN Bus
Based on Deep Features of Voltage Signals
}

\author{\IEEEauthorblockN{Efrat Levy\IEEEauthorrefmark{1} Asaf Shabtai\IEEEauthorrefmark{1} Bogdan Groza\IEEEauthorrefmark{2} Pal-Stefan Murvay\IEEEauthorrefmark{2} Yuval Elovici\IEEEauthorrefmark{1}}
\IEEEauthorblockA{\IEEEauthorrefmark{1}Dept. of Software and Information Systems Engineering, Ben-Gurion University of the Negev \\}
\IEEEauthorblockA{\IEEEauthorrefmark{2}Politehnica University of Timisoara
}
}



\maketitle


\begin{abstract}
The Controller Area Network (CAN) is used for communication between in-vehicle devices.
The CAN bus has been shown to be vulnerable to remote attacks. To harden vehicles against such attacks, vehicle manufacturers have divided in-vehicle networks into sub-networks, logically isolating critical devices. However, attackers may still have physical access to various sub-networks where they can connect a malicious device. This threat has not been adequately addressed, as methods proposed to determine physical intrusion points have shown weak results, emphasizing the need to develop more advanced techniques.
To address this type of threat, we propose a security hardening system for in-vehicle networks. The proposed system includes two mechanisms that process deep features extracted from voltage signals measured on the CAN bus. The first mechanism uses data augmentation and deep learning to detect and locate physical intrusions when the vehicle starts; this mechanism can detect and locate intrusions, even when the connected malicious devices are silent. This mechanism's effectiveness (100\% accuracy) is demonstrated in a wide variety of insertion scenarios on a CAN bus prototype.
The second mechanism is a continuous device authentication mechanism, which is also based on deep learning; this mechanism's robustness (99.8\% accuracy) is demonstrated on a real moving vehicle.
\end{abstract}

\newcommand\setrow[1]{\gdef\rowmac{#1}#1\ignorespaces}
\newcommand\clearrow{\global\let\rowmac\relax}
\clearrow

\begin{figure*}[t!]
\scriptsize
\centering
\begin{minipage}{13 cm}
\includegraphics[width=13 cm]{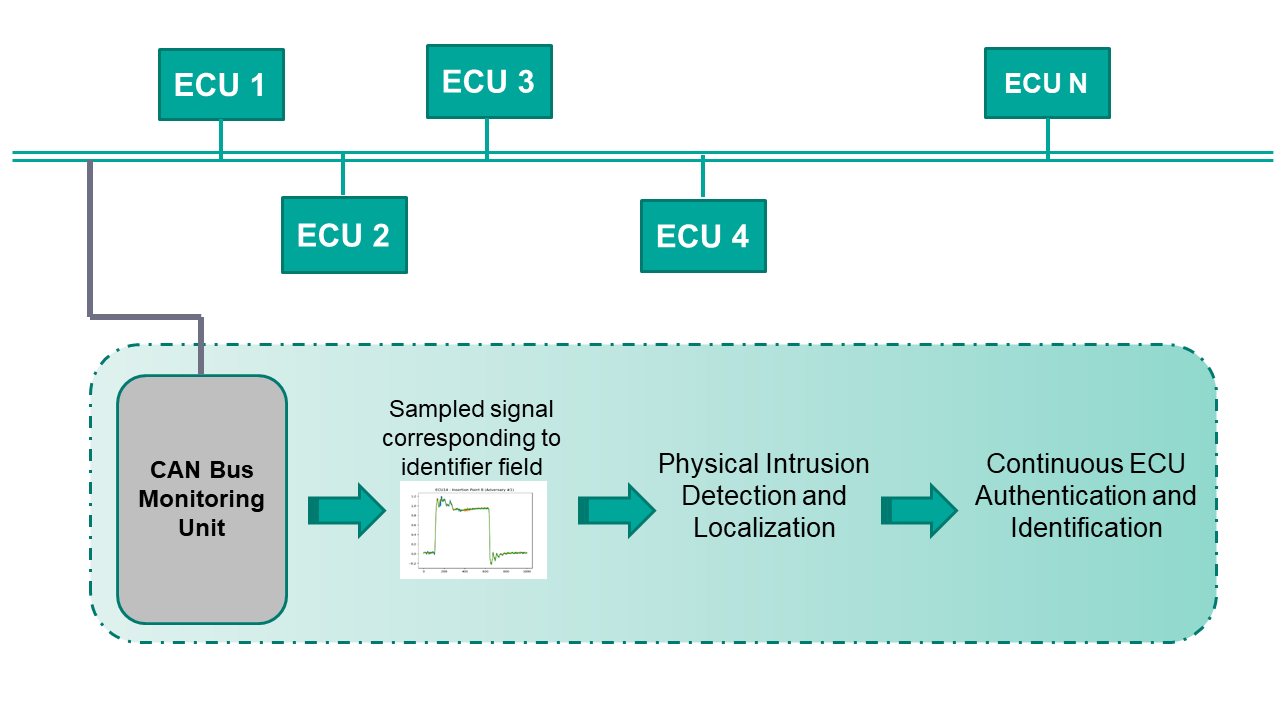}
\centering
\end{minipage}
\caption{Example of CAN bus line topology with a continuous monitoring unit connected to the CAN bus. This unit is responsible for sampling and analyzing voltage signals transferred on the CAN bus.}
\label{fig:can_bus}
\end{figure*}

\section{\label{sec:intro}Introduction}

The Controller Area Network (CAN) protocol has been widely adopted for real-time communication between electronic control units (ECUs) in modern vehicles~\cite{tuohy2014intra}. 
The CAN protocol was designed to provide a high level of fault tolerance, however less attention was paid to security issues (e.g., authentication), which were not a major source of concern when it was developed. 
These unaddressed security issues make the CAN protocol vulnerable to modern threats, such as denial-of-service (DoS) attacks~\cite{woo2014practical} and spoofing~\cite{miller2015remote,palanca2017stealth,liu2017vehicle}.

As a case in point, it has been shown that an attacker can launch a spoofing attack and send falsified frames via a malicious diagnostic tool connected to the on-board diagnostics (OBD) port or a compromised telematic control unit ~\cite{liu2017vehicle,sagong2018cloaking}.
Spooﬁng and DoS attacks targeting in-vehicle ECUs have also been shown to be feasible as well~\cite{miller2015remote,woo2014practical}. 
Since the CAN protocol is the automotive industry standard, the security issues of the CAN bus have become a major concern of vehicle manufacturers and have been the focus of a growing amount of research.
To harden vehicles against remote attackers, vehicle manufacturers have divided in-vehicle networks into sub-networks, logically isolating critical ECUs from the Internet. However, the significant threat from attackers with physical access to the CAN bus remains unaddressed.

In this study, we focus on the security of the CAN bus in two different respects. The first is defending against attackers with physical access to the CAN bus, and the second is defending against spoofing attacks, whether performed remotely or locally (e.g., through a supply chain attack).

There has been very little research attention given to: (1) evaluating the intrusion detection method when the attack involves replacing an existing ECU with a malicious one or connecting a new malicious ECU to the CAN bus, or (2) identifying a malicious ECU's location on the CAN bus; the latter is very important for mitigating potential attacks originating from a malicious ECU added to the CAN bus. The method presented in ~\cite{murvay2020tidal} represents an initial attempt at intrusion point localization, but it is unable to accurately localize the physical intrusion point when new ECUs have been added to the CAN bus.

A common approach for mitigating spoofing attacks on the CAN bus is to add a cryptography-based authentication mechanism~\cite{harel2019optimizing, 6093081}.
However, cryptographic authentication requires that all ECUs support complex cryptographic operations, which consume a lot of memory and computation.
Such an approach raises backward compatibility issues and necessitates demanding key management procedures; moreover, accommodating cryptographic material in the limited 64-bit payload of CAN frames is itself challenging. Consequently, much more work is needed before CAN networks can fully support cryptographic algorithms.
 
There is, however, another means of coping with spoofing attacks that \textit{does not} require any changes to the protocol -- authenticating connected ECUs by analyzing and modeling their communication on the CAN bus.
This can be done by performing a timing analysis of the frames, using various statistical and machine learning-based mechanisms~\cite{murvay2020tidal,moore2017modeling,cho2016fingerprinting,kulandaivel2019canvas} or by conducting payload-based analysis~\cite{woo2016practical,zago2017quantitative}. 
However, research has demonstrated that an attacker can evade detection by such mechanisms~\cite{sagong2018cloaking,murvay2020tidal}; for example, the attack can replicate the propagation delay behavior of a legitimate frame transmitter~\cite{murvay2020tidal}. 

Taking the evasion constraint into consideration, previously proposed methods have used the unique characteristics of voltage signals generated during transmissions by each individual ECU in order to detect spoofing attacks~\cite{yang2020identify,xu2019voltage}.
Compared to the timing-based and payload-based methods, this approach is more difficult to evade.

In a recent study~\cite{bhatia2021evading}, the researchers show a novel technique to evade spoofing detection mechanisms which are based on voltage signals analysis. In their work, they show an exploitation to the detection mechanism retraining process by connecting a malicious ECU to the CAN bus.

In order to secure the CAN bus, we propose \MethodName, a security hardening system for in-vehicle networks, which is based on features derived from voltage signals transferred on the CAN bus. Our proposed system consists of two mechanisms. The first is a physical intrusion detection and localization mechanism, which uses a deep autoencoder that detects changes in the network topology and convolution neural network (CNN) classifiers that report the exact location of malicious insertions or ECU replacements.
The second is a continuous authentication and identification mechanism, which uses CNN classifiers and is capable of detecting spoofing attacks by legitimate ECUs that impersonate their peers. 

From a practical standpoint, the proposed system is comprehensive in that the physical intrusion detection and localization mechanism runs once when the vehicle is started, attempting to detect and locate changes that have been made to the network, and the continuous authentication and identification mechanism runs continuously after the vehicle has been started.

Our system design is inspired by recent power analysis research in which classification using deep learning has been shown to be more powerful and robust than statistical methods ~\cite{timon2019non, das2019x, wegener2019dl}.
In the course of our research, we derived the novel insight that information related to physical intruders and their location is encoded within the legitimate signals' voltage transferred on the CAN bus. Thus, our system is effective against silent ECUs maliciously connected to the CAN bus.

We validate the physical intrusion detection and localization mechanism on a CAN bus prototype using a large dataset of ECU replacement and insertion attacks, and show that our mechanism can detect changes in the network topology with 100\% accuracy and locate physical intrusions with 100\% accuracy for insertion scenarios and 98\% to 100\% for replacement scenarios.

We validate the authentication mechanism on a CAN bus prototype and traffic recorded from a real vehicle, i.e., a 2015 Honda Civic. We demonstrate the robustness of our mechanism under the following demanding conditions: training using data collected while the vehicle is stationary and testing it over a long period of time (over an hour) when the vehicle is moving. Our evaluation results on a real vehicle show 99.8\% ECU identification accuracy when the vehicle is moving (similar results are achieved when the evaluation is performed on a CAN bus prototype).

The main contributions of this study are summarized as follows:
\begin{itemize}
    \item We present a deep learning-based mechanism combined with a data augmentation technique that allows the detection and localization of physical intruders, even when they are silent.
    
    \item We perform a comprehensive evaluation of physical intrusion detection and localization on a CAN bus prototype, using a wide variety of intrusion attacks.
    
    \item We present a deep learning-based mechanism for learning the unique characteristics (patterns) of individual ECUs based on the ECUs' voltage signals measured on the CAN bus. 
    
    \item We perform an evaluation of the authentication mechanism's robustness on both a CAN bus prototype and a real vehicle when moving.
    
    \item This research complements a recent study presenting a prevention solution which requires accurate localization capability~\cite{canary2021bogdan}.
\end{itemize}

\section{\label{sec:background}Background}

\subsection{CAN Communication}

\begin{figure}[th!]
\includegraphics[width=0.48\textwidth]{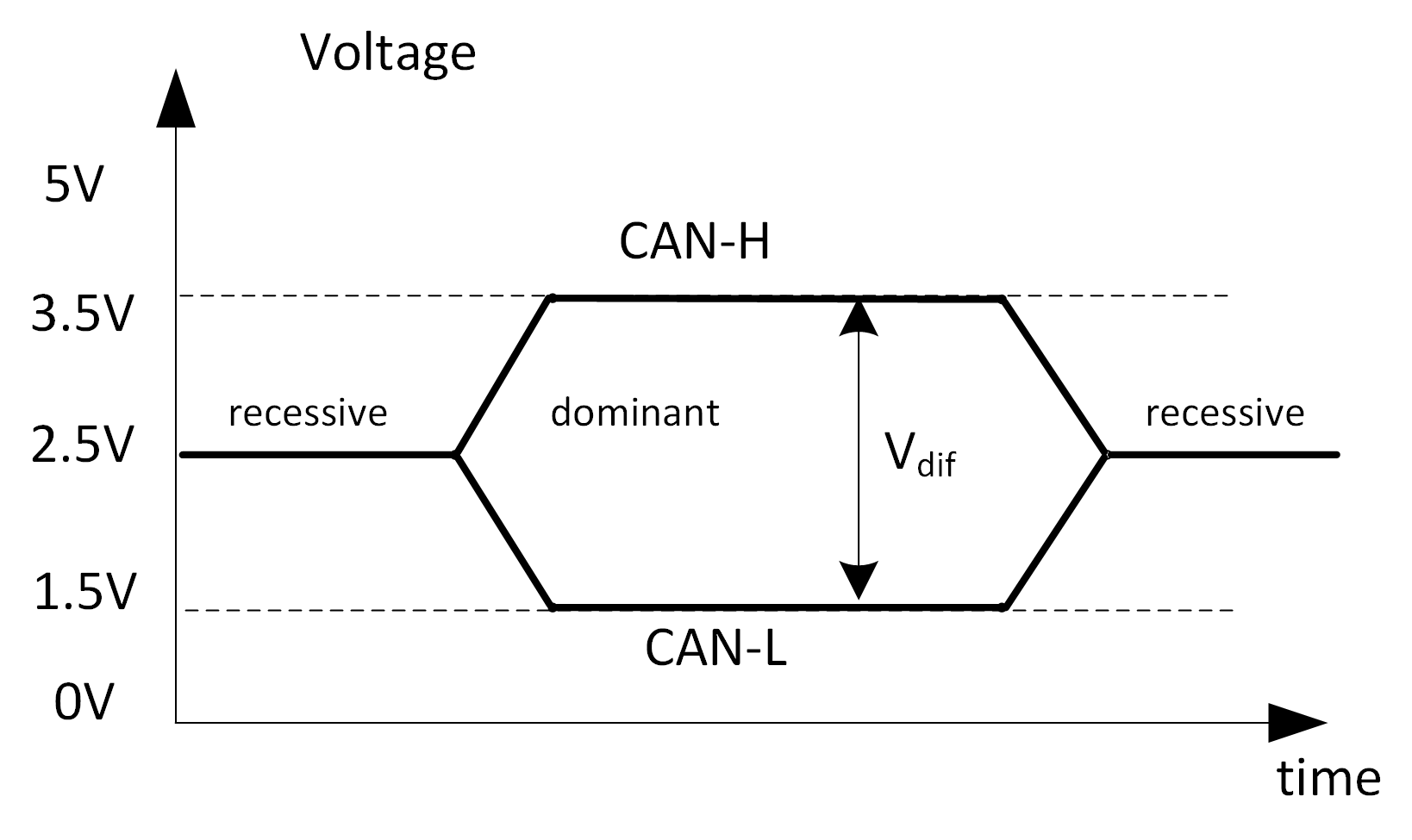}
\caption{Nominal voltage of the CAN-H and CAN-L lines during the recessive and dominant states.}
\label{fig:can_phy}
\end{figure}

The CAN bus is a two-wire broadcast bus. It uses the differential voltage between the two bus lines, CAN-H and CAN-L, in order to encode the bits. During the dominant state, the CAN-H line is driven toward a nominal voltage of 3.5V, and the CAN-L line is driven toward a nominal voltage of 1.5V. The resulting differential voltage $V_{\mathit{diff}}$ during the dominant state must be within 0.9-2.0V, a case in which a ``0" is interpreted by the ECU transceiver. For the recessive state, both the CAN-H and CAN-L lines are driven toward a nominal voltage of 2.5V, and a ``1" is interpreted for a differential voltage less than 0.5V. An illustration of the differential voltage is presented in Figure \ref{fig:can_phy}.

\begin{figure}[th!]
\includegraphics[width=0.47\textwidth]{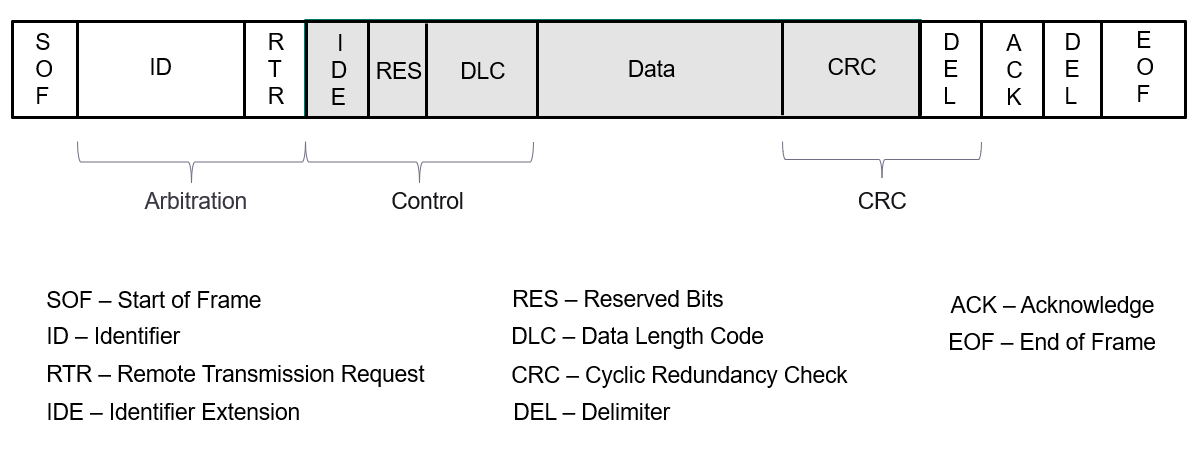}
\caption{Structure of a standard CAN frame.}
\label{fig:can_frame}
\end{figure}

Figure \ref{fig:can_frame} presents a standard CAN frame structure. 
The CAN frame begins with the start of frame (SOF) bit, which is a ''0" bit that drives the CAN bus from the recessive state to a dominant state. The identifier field ID, which is used in the arbitration, is next. Since multiple ECUs can write on the CAN bus at the same time, an arbitration mechanism is needed to avoid collisions. The arbitration mechanism is based on the message identifier (ID), which is the first field after the start of frame (SOF). Lower valued identifiers have the highest priority; note that dominant bits, i.e., zeros, will always overwrite recessive bits, i.e., ones. Several control fields follow the identifier field ID: the RTR bit, which signals remote frames; the IDE bit, which signals the extended identifier; a reserved field, which signals future extensions; and the DLC field, which represents the length of the data field. The latter, which represents the actual data, can occupy up to eight bytes.
This field is followed by a 15-bit CRC and a delimiter. The acknowledgement field, ACK, is written by all ECUs that successfully receive the frame. It is followed by a delimiter and the end-of-frame (EOF).

Since multiple ECUs can write on the CAN bus at the same time, an arbitration mechanism is needed to avoid collisions. The arbitration mechanism is based on the message identifier (ID), which is the first field after the start of frame (SOF). Lower valued identifiers have the highest priority; note that dominant bits, i.e., zeros, will always overwrite recessive bits, i.e., ones.

\subsection{ECU Voltage Signals} \label{voltage_signals}

\begin{figure}[th!]
\centering
\includegraphics[width=0.45\textwidth]{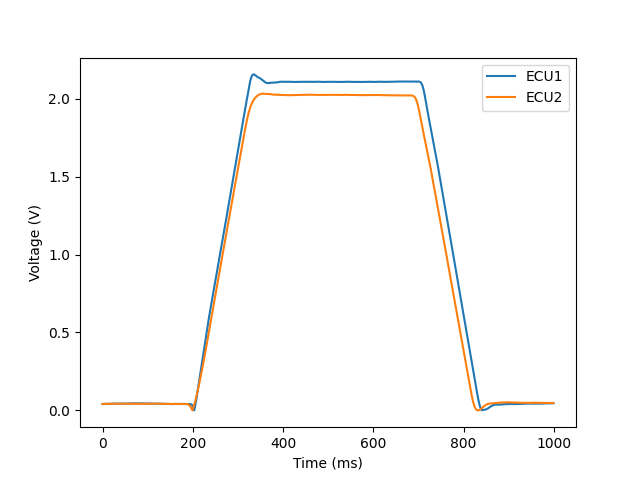}
\caption{ A demonstration of a recessive ''1" to a dominant
''0" transition and return (differential voltage $V_{\mathit{diff}}$ recorded
from two distinct ECUs).}
\label{fig:analog_signals}
\end{figure}

A modern vehicle contains a variety of ECUs. Each ECU generates unique analog signals. Even if the same CAN frames are transmitted by two identical ECUs manufactured in the same batch, their signals' characteristics are different. Recent studies showed that these characteristics are useful for highly accurate ECU fingerprinting ~\cite{gerdes2012physical,hafeez2019ecu}. 
When analyzing the digital representation of a sampled signal, those differences are expressed by relatively minor changes. Figure \ref{fig:analog_signals} visually illustrates the difference between the signals of two ECUs, as sampled from the rising and falling edges of a CAN frame. 

Each CAN bus ECU outputs a signal that has unique physical characteristics which are due to both manufacturer specific designs and tiny imperfections in the components, e.g., the ECU's transceiver's internal resistance and capacitance. Furthermore, each ECU added to the CAN bus contributes its own resistance and capacitance, modifying the overall electronic characteristics of the CAN bus and thus affecting the signals of all existing ECUs. The influence of this differs according to the connection location and the ECUs' transceiver characteristics.

\begin{figure}[th!]
\centering
\includegraphics[width=0.45\textwidth]{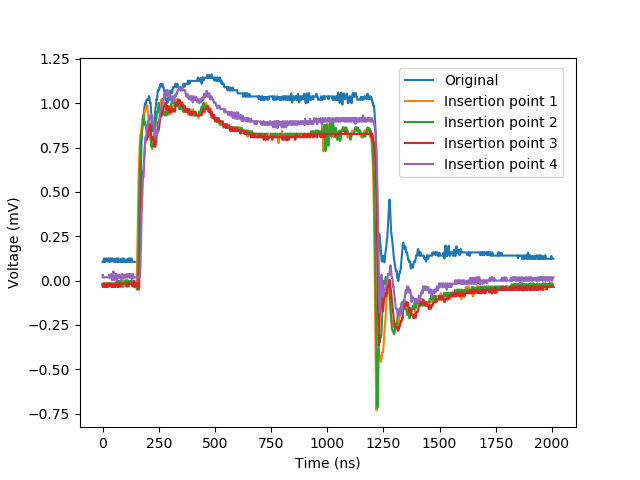}
\caption{A demonstration of how existing ECU signals are influenced when an ECU is added at different locations on the CAN bus. }
\label{fig:influence}
\end{figure}

\begin{figure}[th!]
\centering
\includegraphics[width=0.45\textwidth]{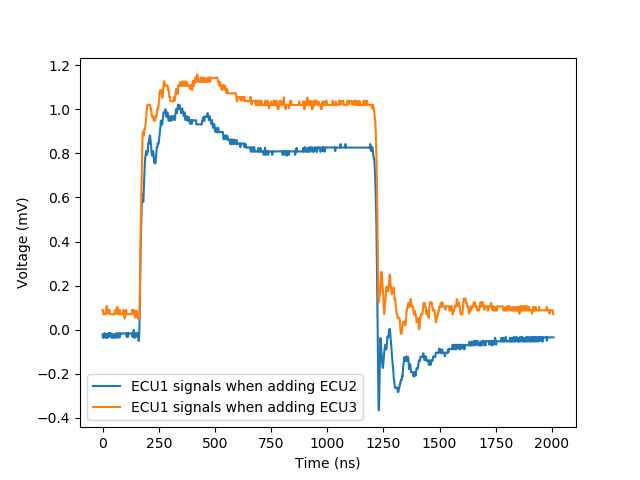}
\caption{A demonstration of how existing ECU signals are influenced when two different ECUs are added at the same location on the CAN bus. }
\label{fig:module_influence}
\end{figure}

Figure \ref{fig:influence} illustrates the changes to the existing ECU signals when a new ECU is introduced at various insertion locations on the CAN bus.  
Figure \ref{fig:module_influence} illustrates the changes to existing ECU signals when two ECUs from different manufacturers are introduced at the same location on the CAN bus.

\begin{table*}[t]
\caption{Summary of related work} \label{tab:relatedworks}
\centering
\scriptsize
\begin{tabular}{|c|c|c|c|c|p{0.135\linewidth}|p{0.15\linewidth}|p{0.12\linewidth}|c|}
\hline
\multirow{2}{*}{Ref.} & \multicolumn{3}{c|}{Attack vector}                                                                                                                                         & \multirow{2}{*}{\begin{tabular}[c]{@{}c@{}}Intrusion point \\ localization\end{tabular}} & \multirow{2}{*}{Method used} & \multirow{2}{*}{Features} & \multirow{2}{*}{Experimental testbed} & \multirow{2}{*}{\begin{tabular}[c]{@{}c@{}}Sampling\\ frequency\end{tabular}} \\ \cline{2-4}
                      & \begin{tabular}[c]{@{}c@{}}Compromise\\ ECU\end{tabular} & \begin{tabular}[c]{@{}c@{}}Add new \\ ECU\end{tabular} & \begin{tabular}[c]{@{}c@{}}Replace \\ ECU\end{tabular} &                                                                                           &                              &                           &                                       &                                                                               \\ \hline
\cite{murvay2014source} & \checkmark & - & - & Not relevant & Signal processing & Raw signal & CAN bus prototype & 2 GS/s \\
\hline

\cite{cho2017viden} & \checkmark & - & - & Not relevant & Signal processing & Statistical features extracted from the raw signal & CAN bus prototype \& two real cars & 50 kS/s \\
\hline

\cite{choi2018identifying} & \checkmark & \checkmark & - & - & ML (SVM, NN, BDT) & 9 frequency domain \& 8 time domain features & CAN bus prototype & 2.5 GS/s \\
\hline

\cite{choi2018voltageids} & \checkmark & - & - & Not relevant & ML (LiSVM, BDT) & 9 frequency domain \& 8 time domain features & CAN bus prototype & 2.5 GS/s \\
\hline

\cite{kneib2018scission} & \checkmark & \checkmark & - & - & ML (logistic regression) & Features extracted from rising and falling edges & CAN bus prototype \& two real cars & 20 MS/s \\
\hline

\cite{kneib2019robustness} & \checkmark & \checkmark & - & - & Signal processing & Features extracted from rising and falling edges & One real car & 2 MS/s \\
\hline

\cite{foruhandeh2019simple} & \checkmark & - & - & Not relevant & Statistical analysis & Temperature and voltage & CAN bus prototype & 50 MS/s \\
\hline

\cite{rumez2019can} & - & \checkmark & - & Additions only & Statistical analysis & Response to sent pulses & CAN bus prototype & 2 GS/s$^2$ \\
\hline

\cite{xu2019voltage} & \checkmark & - & - & Not relevant & Reinforcement learning  & Raw signal (sampled from a dominant (0) bit) & CAN bus prototype & N/A \\
\hline

\cite{yang2020identify} & \checkmark & - & - & Not relevant & ML (deep learning) & Raw signal & CAN bus simulation  & 250 MS/s \\
\hline

\cite{hafeez2019ecu} & \checkmark & - & - & Not relevant & ML (deep learning) & Statistical features extracted from the raw signal & CAN bus prototype & 2 GS/s \\
\hline

\hline
\hline
\rowcolor[gray]{0.9}
\MethodName & \checkmark & \checkmark & \checkmark & \checkmark & ML (deep learning) & Raw signal sampled from rising and falling edges & CAN bus prototype \& one real car & 500 MS/s \\
\hline

\end{tabular}
\end{table*}

\section{\label{sec:related}Related Work}

In contrast to timing-based or payload-based analysis, our mechanism relies on CAN bus electrical signals, which are difficult to fake.
Therefore, as related work, we only consider physical intrusion and spoofing detection mechanisms that are based on features extracted from electrical signals.

In previous studies, several methods to detect spoofing attacks based on ECUs' electrical signals were proposed. Table ~\ref{tab:relatedworks} summarizes and compares this research based on the following criteria: attack vector, intrusion point localization, detection methods used, extracted features, experimental testbed setup, and signal sampling frequencies.

The first study presenting the idea of using voltage signals for ECU fingerprinting used simple signal processing techniques that were applied on the raw signal sampled from the CAN frame's arbitration field ~\cite{murvay2014source}.
Another study ~\cite{cho2017viden} proposed adaptive signal processing applied on statistical features extracted from the raw signal; the proposed mechanism enables modification of the fingerprints and hence allows the mechanism to adapt to possible environmental changes. 

In other research presented by Choi et. al.  ~\cite{choi2018identifying}, the authors presented improvements related to signal processing. In this study, 17 features were extracted from the extended identifiers, and a variety of machine learning algorithms were employed in order to improve the identification accuracy obtained in prior work. 

Further improvements were presented in a subsequent study ~\cite{kneib2018scission} in which higher accuracy was achieved although simpler machine learning algorithms were used. The core idea behind that study is the observation that the identification accuracy can be significantly improved by processing the samples of the rising and falling edges of the transmitted signals over the CAN bus; the samples are acquired after the arbitration field. Other research presented an improvement in terms of the sampling frequency used during data collection ~\cite{kneib2020easi}.

Choi et. al ~\cite{choi2018voltageids} have presented additional improvements. Similarly to ~\cite{kneib2018scission}, they sampled the rising and falling edges of the transmitted signals while using the same feature extraction presented in ~\cite{choi2018identifying}.

Significant improvements in terms of computational and data collection resources were achieved in another study ~\cite{foruhandeh2019simple} in which statistical analysis was applied on either temperature or voltage variations, serving as an adaptive approach for addressing environmental changes.

One property shared by the studies presented above is that they are all passive. A different approach was taken in a study ~\cite{rumez2019can} that used time-domain reflectometry (TDR), in which a pulse is sent on the CAN bus, and the response is measured. While this technique can locate the connected ECUs on the CAN bus and detect changes in the CAN bus topology, it has two significant drawbacks in contrast to our approach: (1) it does not allow ECU fingerprinting (thus, it cannot detect replacement scenarios or spoofing scenarios), and (2) it is an active technique that depends on an active operation on the CAN bus to detect topology changes.

In another line of research, optimization of the passive authentication techniques mentioned above was suggested ~\cite{xu2019voltage}. This approach is based on reinforcement learning, which allows authentication optimization via a trial and error mechanism without prior knowledge regarding the signal or spooﬁng model.

More recently, deep learning techniques have been suggested ~\cite{yang2020identify}. This study used an RNN-LSTM multiclass classifier for the authentication of ECUs on the CAN bus given a raw voltage signal. In other research, a combination of feature extraction and a deep learning-based mechanism was suggested ~\cite{hafeez2019ecu}.
Both methods achieved good identification accuracy, however the proposed models' robustness to environmental changes was not demonstrated; the studies also did not address the detection and localization of a malicious ECU device connected to the CAN bus.

In order to address the limitations of the prior work mentioned above, in this work, we propose a robust system which focuses on the security of the CAN bus in two different respects. The first is defending against attackers with physical access to the CAN bus and the second is defending against spoofing attacks, whether performed remotely or locally (e.g., through a supply chain attack).

Based on the \textbf{legitimate} ECUs' signals transferred on the CAN bus, our proposed system determines whether the CAN bus has been physically modified. To ensure driver safety, this process is executed when the vehicle is started. Our system takes advantage of the fact that each CAN bus topology change influences all of the voltage signals transferred on the CAN bus. Moreover, we show that those legitimate signals are also useful for locating the physical intruder on the CAN bus; we derived the novel insight that the intruder's location is encoded within the legitimate ECUs' signals. Thus, our mechanism can detect and locate silent ECUs introduced at an available location of the CAN bus.

In addition, we propose a robust ECU authentication mechanism that allows the detection of spoofing attempts continuously after the vehicle has been started. Regarding intrusion localization scenarios, thanks to the ability of the proposed authentication mechanism to generalize, we show that the proposed authentication mechanism is also useful for locating physical intruders when a legitimate ECU is replaced. This is done by applying a process of monitoring (and identifying) legitimate signals until the missing ECU which was replaced is detected.

In order to evaluate our proposed system, we used a CAN bus prototype identical to that of ~\cite{murvay2020tidal}, which used timing analysis in order to authenticate and locate a malicious device connected to the CAN bus. The current study differs in the following ways. First, in~\cite{murvay2020tidal}, only the difference in the arrival time was used (extracted by setting a threshold for the voltage level); the shape of the signal on the CAN bus is ignored. In our study, we show how deep learning techniques can be used to delve further into specific patterns of the voltage signal that are unique to each ECU. Second, their method requires a connection to each end of the CAN bus, whereas our method only requires one connection to the CAN bus, which simplifies the wiring harness. Third, their method was unable to localize the physical intruder in cases in which a new ECU was inserted into the CAN bus, i.e., a change in the voltage characteristics of the CAN bus. By using deep learning with data augmentation, we can localize malicious ECUs, even when they are unknown to the mechanism and/or silent.

In order to demonstrate the robustness of our authentication mechanism compared to approaches proposed in prior studies, we used voltage signals collected from both a CAN bus prototype and a real vehicle. The current study differs from prior research in the following ways. First, by using a variety of insertion scenarios on a CAN bus prototype, we show that our authentication mechanism is robust to CAN bus topology changes. Second, we generate the ECU fingerprints using a relatively small amount (a few thousands) of voltage signals collected while the vehicle is stationary and test it for a long period of time (over an hour) when the vehicle is moving.

\section{\label{sec:threatmodel}Network and Threat Model}

\subsection{Network Model}
Our proposed detection system requires physical access to the network. While in-vehicle networks may have more than a hundred ECUs, they are always grouped together in sub-networks of less than a dozen ECUs.

The typical sub-network topology is bus oriented. In this topology, a two-wire cable connects multiple ECUs that implement various car functionalities, as illustrated in Figure \ref{fig:can_bus}. To protect the entire vehicle, our system must be connected to each sub-network in order to sample signals from each of the existing buses. Alternatively, our system can be deployed on critical sub-networks only.

\subsection{Threat Model}

We consider two types of attackers:
\begin{itemize}
    \item An attacker with physical access to the CAN bus, aiming to replace an existing ECU with a malicious device or insert an additional device at a specific location.
    \item A remote attacker that exploits a vulnerable device, aiming to write spoofed messages on the CAN bus and take control of critical sub-systems.
\end{itemize}

\begin{figure}[h]
\includegraphics[width=0.47\textwidth]{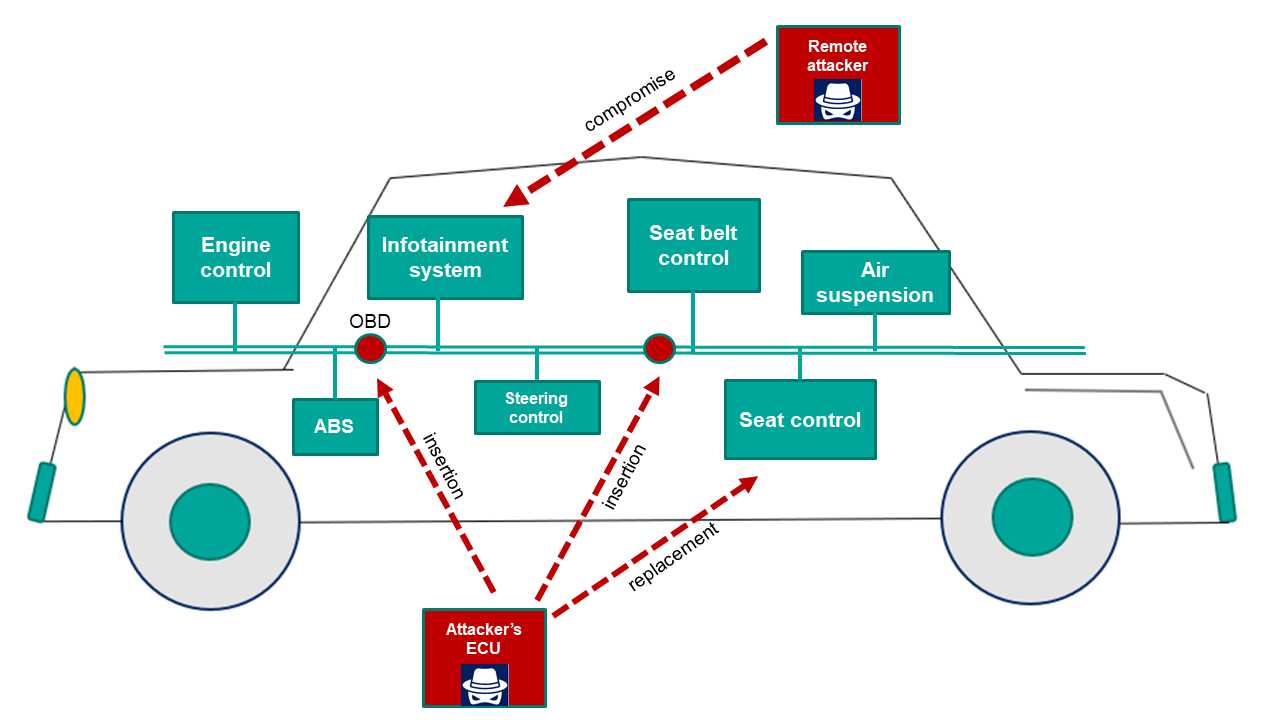}
\caption{Surfaces that can be used to conduct attacks on the CAN bus.}
\label{fig:attack_surfaces}
\end{figure}

An illustration of the attack surfaces is presented in Figure \ref{fig:attack_surfaces}. These include both open entry points to the CAN bus (e.g., the OBD port), as well as existing ECUs that can be corrupted (e.g., infotainment systems), and other bus taps that can be installed by an adversary in accessible locations.

We assume that an attacker is aware of the presence of the detection mechanism and how it works. The attacker can obtain this information by reverse engineering or inside information. Therefore, our proposed method is based on analyzing the hardware's unique physical characteristics which makes evasion infeasible.

\section{\label{sec:highlevel} High-Level Description of the System}

\begin{figure}[h]
\centering
\includegraphics[height=0.47\textwidth]{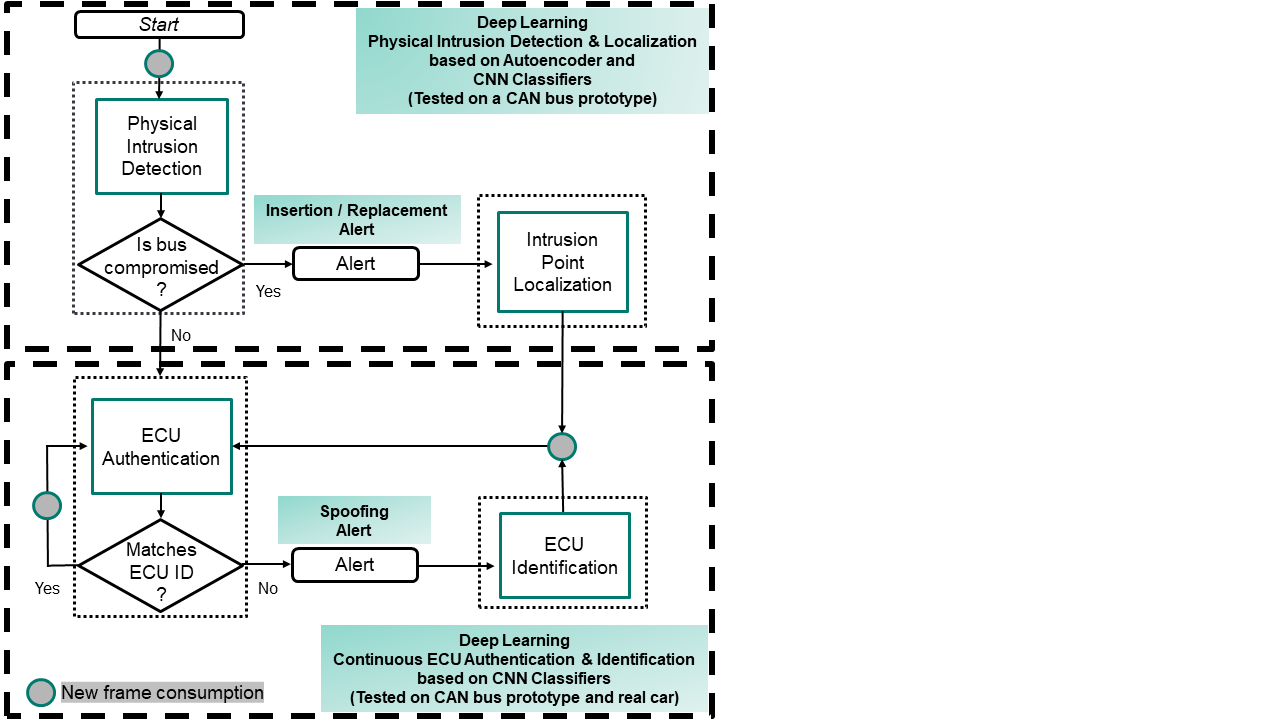}
\caption{High-level architecture of the proposed system.}
\label{fig:high_level}
\end{figure}

In order to secure the CAN bus from physical intruders and remote/local spoofing attacks, we propose a system which is based on continuous monitoring of the analog signals transferred on the CAN bus. The proposed system (illustrated in Figure \ref{fig:high_level}) consists of two mechanisms: 
\begin{enumerate}
    \item Physical Intrusion Detection and Localization - this mechanism is active once when the vehicle is started. It detects CAN bus network topology changes and locates physical intruders.
    \item Continuous ECU Authentication and Identification - this mechanism runs continuously after the vehicle has been started. It detects spoofing attempts and identifies the real origin of the spoofed message.
\end{enumerate}

The advantages of the proposed system are twofold. First, the entire system is based on analyzing voltage signals transferred on the CAN bus. We claim that such signals are uniquely generated by each device due to hardware inconsistencies. Thus, our system is more robust to detection evasion related to other detection solutions like timing-based and payload-based solutions. Second, for intrusion detection and localization, our method does not depend on signals transferred by the intruder device. Our method only analyzes known ECUs' signals, and based on these signals, it is able to detect and locate new intruders.

A new ECU can be introduced by inserting a new ECU into an available location on the CAN bus or by replacing an existing ECU. Other cases (e.g., swapping the locations of two existing legitimate ECUs) are not considered in this work.

\subsection{Data Acquisition}
When data is acquired from the CAN bus, \MethodName samples the physical signal of CAN frames. Each CAN frame can be associated with a particular sender based on its ID field. While the CAN bus is a broadcast bus, in existing practical implementations, each ECU is associated with a set of IDs that it uses to send data on the CAN bus. Remote frames (which request specific data) with the same ID as data frames can be sent by distinct ECUs, but since this type of frame does not carry any data, it cannot be a source of an impersonation attack and is not relevant to our analysis. Remote frames are easily distinguished by the RTR bit, which is set at one. 

One of our goals is to authenticate each of the legitimate ECUs based on the sampled signals. Therefore, when generating the fingerprints we need to associate each sampled signal with its ECU. Since other ECUs are allowed to transmit information in the arbitration and acknowledgement fields, the only fields that can be sampled for ECU identification are the control, data, and CRC fields (gray ﬁelds in Figure \ref{fig:can_frame}). As shown in previous studies ~\cite{kneib2018scission}, rising or falling edges should be sampled in order to increase the detection accuracy.

In this study, we assume that the original topology of the CAN bus is known to the system. In particular, the locations of the legitimate ECUs are known. 

\subsection{Proposed System Description}
\textbf{Physical Intrusion Detection and Localization.} This mechanism is responsible for detecting changes in the network topology of the CAN bus. In particular, it determines whether the CAN bus is \emph{clean} (no ECU was replaced or added to the CAN bus) or \emph{dirty} (a new ECU was added or replaced an existing ECU), i.e., the CAN bus is compromised. If the CAN bus is compromised, an alert is generated, and the intrusion point location is returned. As illustrated in Figure \ref{fig:high_level}, two modules are proposed: (i) the physical intrusion detection module, and (ii) the intrusion point localization module. 

Algorithm 1 describes the physical intrusion detection and localization mechanism. The input to the algorithm is the inspected signal (denoted by $sig$), which is a list of voltage samples collected from the CAN bus during a frame transmission. First, the physical intrusion detection module is used to detect whether the CAN bus is compromised (line 2). If the CAN bus is compromised, an alert is generated (line 3), and then the physical intrusion localization module is used to locate the physical intrusion point (line 4).

The physical intrusion localization module is based on a process of monitoring legitimate ECUs signals (line 6). The main building block of the monitoring process is the authentication mechanism, which is applied on the transmitted frames when the CAN bus is compromised. This process is performed to distinguish between insertion and replacement scenarios: 
\begin{itemize}
    \item All known ECUs have been identified within a given time period (line 7). In this case, we conclude that it is an insertion attack, and an insertion localization procedure is executed (line 8) to return the insertion location (line 9).
    \item If the time period has ended, and there is a known ECU that has not been identified (line 10), we conclude that it is a replacement attack. In this case, the location of the missing ECU is returned (line 11).
\end{itemize}

\begin{algorithm}
\scriptsize
\caption{Physical Intrusion Detection \& Localization}
\begin{algorithmic}[1]
\Procedure{DetectPhysicalIntrusion}{$sig$} 
        \If{$IsBusCompromised(sig)$}
           \State  $GenerateAlert()$
            \State \textbf{return} $LocatePhysicalIntrusion()$
        \EndIf
\EndProcedure

\Procedure{LocatePhysicalIntrusion}{} 
        \State $M \leftarrow Monitor.getMissingECUs()$
        \If{$M = \emptyset $}
            \State $S \leftarrow Monitor.getMonitoredSignals()$
            \State $location \leftarrow LocateInsertionPoint(S)$
        \Else
            \State $location \leftarrow LocateReplacementPoint(M)$
        \EndIf
        \State \textbf{return} $location$
\EndProcedure
\label{proposed_alg_phase1}
\end{algorithmic}
\end{algorithm}

In this study, we assume that in-vehicle ECUs transmit frames periodically, although the presence of a silent ECU is technically possible. However, this would be uncommon, since each ECU handles several functionalities and must periodically report data from various sensors/actuators.

\textbf{Continuous ECU Authentication and Identification.} This mechanism is responsible for continuously detecting spoofing attempts. In this case, an alert is generated, and the real origin of the spoofed message is returned. As illustrated in Figure \ref{fig:high_level}, two modules are proposed: (i) the ECU authentication module, and (ii) the ECU identification module. 

The input to the module is the inspected signal, which is a list of voltage samples collected from the CAN bus during a frame transmission and the identifier of the ECU transmitting it. First, the ECU authentication module is used to authenticate the frame given the voltage signal and the claimed identifier. If there is no match, an alert is generated, and then the ECU identification module is used to return the real sender of the frame.

\section{\label{sec:proposedmethod} Low-Level Description of the System}

In this section, we provide a detailed description of the proposed system.

\subsection{Physical Intrusion Detection}
The physical intrusion detection module is responsible for detecting changes in the network topology of the CAN bus. In particular, it determines whether the CAN bus is \emph{clean} (no ECU was replaced or added to the CAN bus) or \emph{dirty} (a new ECU was added or replaced an existing ECU), i.e., the CAN bus is compromised.

The physical intrusion detection module is implemented by an autoencoder which receives a voltage signal that is transferred on the CAN bus and determines whether the CAN bus is compromised. An autoencoder is an unsupervised algorithm that represents input data in a lower dimensionality and then reconstructs the data to its original dimensionality; thus, the normal instances are reconstructed properly, and the outliers are not. In this way, anomalous input data can be identified.

As described earlier in Section \ref{voltage_signals}, the basis for this module is the electric property of CAN bus topologies, in which each network topology change influences all of the signals transferred on the CAN bus. Since any new ECU that taps the CAN bus affects the voltage signals of all of the ECUs, a single CAN frame (regardless of the sender) is sufficient for detecting whether the CAN bus topology has changed.

\textbf{Autoencoder architecture.} The autoencoder consists of two parts: the encoder and the decoder. The encoder learns how to interpret the input and compresses it to an internal representation. 
The decoder takes the output of the encoder and attempts to reconstruct the input. 
We define the encoder so it has two hidden layers
set at decreased sizes of 50 percent and 25 percent of the input layer's dimension.
To ensure that the model learns well, we use batch normalization and leaky ReLU activation. The decoder is defined with a similar structure, although in reverse.

\textbf{Training set.} 
The voltage signals transferred on the CAN bus when the network is \emph{clean}. 

\textbf{Training phase.} During the training phase, we use two separate chronological datasets that only contain benign data (i.e., voltage signals transferred on the CAN bus when the network is \emph{clean}), from which the autoencoder learns the patterns of the original CAN bus topology. 

The first dataset is the training set ($TR_{clean}$), and the second dataset is the validation set ($VAL_{clean}$). Given $TR_{clean}$, we train the autoencoder until the \emph{mean squared error} (MSE) reaches its minimum on $VAL_{clean}$.
We use the Adam optimizer and a learning rate of 0.001.
Once the model training is complete, a threshold ($thr$) is determined to discriminate between benign (i.e., voltage signals transferred on the CAN bus when the CAN bus is \emph{clean}) and malicious signals (i.e., voltage signals transferred on the CAN bus when the latter is \emph{dirty}).

The threshold ($thr$) is calculated as the sum of the samples' mean and the standard deviation of the MSE on $VAL_{clean}$:

\begin{equation}
	thr = mean(MSE_{VAL_{clean}}) + std(MSE_{VAL_{clean}}) 
\end{equation}

\textbf{Intrusion detection phase.} Given a voltage signal transferred on the CAN bus, we execute the autoencoder and measure the reconstruction error of the signal. If the reconstruction error exceeds $thr$, an alert is generated, and the intrusion point localization module is used to locate the intrusion point on the CAN bus.

\subsection{Intrusion Point Localization}
The intrusion point localization module is responsible for physically locating the intrusion point on the CAN bus when the latter is \emph{dirty}. First, we need to eliminate a case in which the CAN bus is \emph{dirty} due to the replacement of a legitimate ECU. We identify the replacement of an ECU by monitoring the CAN bus for a certain period of time $TP$, in order to determine whether all of the ECUs are present. 

To do so, an authentication method is proposed. The authentication method proposed for this module is identical to the method described in Section \ref{authentication}. 
Note that this is a case in which the ECUs are being authenticated while their corresponding voltage signals are influenced by an intruder device.

\begin{figure}[h]
\centering
\includegraphics[width=0.48\textwidth]{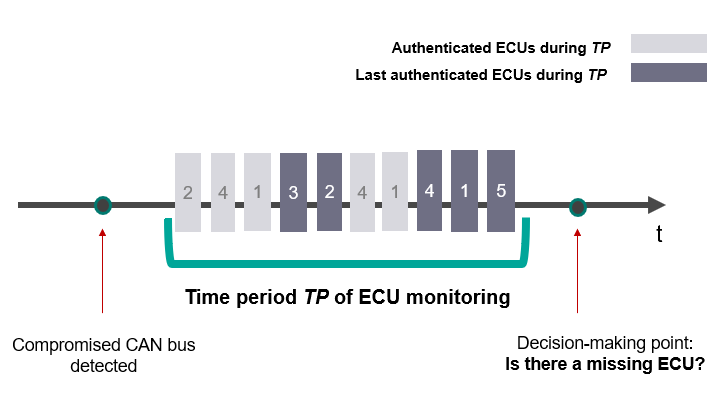}
\caption{Authenticated signal monitoring process on a CAN bus containing five legitimate ECUs.}
\label{fig:monitoring_illustration}
\end{figure}

As illustrated in Figure \ref{fig:monitoring_illustration}, when the vehicle starts, the monitoring process collects authenticated signals during time period $TP$ (one per ECU, and only the last authentication is stored). If all of the ECUs have been successfully authenticated during time period $TP$, the insertion point localization module is used to locate the intruder. Otherwise, the location of the missing ECU is returned.

In order to localize the malicious ECU, one CAN frame from each legitimate ECU needs to be collected. Given the cyclical nature of in-vehicle traffic in which there are predefined cycles (usually in the range of 10-100 ms) and each ECU is in charge of multiple such frames, a few dozen milliseconds, on average, should be sufficient to collect the minimum number of frames and localize the intrusion.

In order to physically locate the intruder in insertion scenarios, a multiclass CNN classifier is proposed.

\textbf{CNN multiclass classifier architecture.}
The proposed architecture is a 1-dimensional variant of VGG16~\cite{simonyan2014very} in which a softmax output layer is attached, providing a probability distribution over the predicted output classes.

Let $P=\{p_1, p_2, ..., p_n\}$ be a set of insertion points on the CAN bus. These points are represented by the classes of the model's output layer. 

\textbf{Training set.} The transmitted signals are collected for a predefined time period at each point $p \in \{p_1, p_2, ..., p_n\}$ when a new ECU is inserted. 
The transmitted signals collected in each time period are labeled with insertion point $p$. During this phase, only signals that are associated with legitimate ECUs are considered.

As demonstrated in Section~\ref{sec:background}, when inserted into the CAN bus, different ECUs influence the transmitted signals differently. Therefore, in order to train a model that estimates the location in general cases, we suggest employing a data augmentation technique. 

Data augmentation is the creation of data from original data, typically by applying a transformation to the original data. Data augmentation is commonly used to improve the versatility of machine learning models, as well as to provide more training examples for datasets of a limited size.
In signal data, for example, it is common to use data augmentation techniques like Gaussian noise addition, cyclic rolling-off (shifting), clipping distortion, and frequency masking~\cite{zhang2019novel, huang2019data}.

Given the changes to existing ECU signals when two ECUs from different manufacturers are introduced at the same location on the CAN bus, adding synthetic data to the training set helped us induce a model that generalizes better and is more accurate. Specifically, given a basic set of signal examples, we extend the set by using the following data augmentation techniques: (1) Gaussian noise addition, and (2) cyclic rolling-off (shifting).

The proposed data augmentation process is described in Algorithm 2. The input to the algorithm consists of the collected signals associated with ECU \emph{i} (denoted by $S^i$). Other input to the algorithm is a set of discrete insertion points (denoted by $P$) and two integers $K$ and $R$. For each insertion point $p \in P$ (line 5) for each signal $s \in S^i_p$ (line 7), we generate $K$ copies of the signal $s$ (line 8). To each copy (line 9), we first add Gaussian noise that is distributed with mean $\mu=0$ and standard deviation $\sigma=1$ (lines 10-11) and then apply a rolling-off (shifting) of a random amount (line 12) of steps. Finally, we assign class $p$ to each signal generated in this loop (line 13). 

\textbf{Training phase.} During the training phase, we use the root mean square propagation (RMSProp) optimizer, with a learning rate of 0.00001., and \emph{categorical cross-entropy} is used as the loss function.
First, we chronologically extract 30\% of the training set to serve as the validation set. Then, we train the network until the loss function reaches its minimum on the validation set.

\textbf{Intrusion localization phase.} As illustrated in Figure \ref{fig:monitoring_illustration}, for each legitimate ECU, the most recent authenticated signal during time period $TP$ is stored. These signals serve as the input to the multiclass classifier in order to locate the intrusion point.

Given $m$ represents the number of legitimate ECUs that are connected to the CAN bus, let $s=\{s_1, s_2,...s_m\}$ be a set of signal vectors (one per ECU). Let $P$ be a matrix such that the column $P_i$ is the multiclass classifier prediction given the input $s_i$. As previously mentioned, $P_i$ represents the probability distribution over the classes (insertion locations on the CAN bus). 

The location estimation technique is presented in Algorithm 3. The input to the algorithm is a group of $m$ signals where signal \emph{i} is associated with legitimate ECU \emph{i} (the group is denoted by $S$). First, we call the multiclass classifier and obtain $|P|=m$ predictions (line 2). Then, we take the most probable class from each column $P_i$ as a class candidate (lines 4-5) and apply a majority over the candidates (line 6). If one candidate remains (line 7), it is returned (line 8). Otherwise, a randomized candidate is returned (line 10). 

\begin{algorithm}
\scriptsize
\caption{Generate Augmented Signals}
\begin{algorithmic}[1]
\Procedure{GenerateSignals}{$S^i,P,K,R$}       
\State $AS^i \leftarrow \emptyset$
\State $\mu \leftarrow 0$
\State $\sigma \leftarrow 1$
    \For {$p \in P$}
        \State $AS^i_p \leftarrow \emptyset$
        \ForEach {$s \in S^i_p$}
            \State $C \leftarrow GenerateCopies(K, s)$
            \For {$c \in C$}
                \State $n \leftarrow RandomizeGaussian(\mu, \sigma)$
                \State $c' \leftarrow AddNoise(n, c)$ 
                \State $r \leftarrow RandomizeUniform([0,R])$ 
                \State $AS^i_p \leftarrow RollOff(r, c')$
            \EndFor
        \EndFor
        \State $AS^i \leftarrow AS^i_p$
    \EndFor
    \State return $AS^i$

\EndProcedure

\label{proposed_alg3}
\end{algorithmic}
\end{algorithm}

\begin{algorithm}
\scriptsize
\caption{Locate Insertion Point}
\begin{algorithmic}[1]
\Procedure{LocateInsertionPoint}{$S$} 
        \State $P \leftarrow Classifier(S)$
        \State $C \leftarrow \emptyset$
        \For {$P_i \in P$}
            \State $C.add(argmax(P_i))$
        \EndFor
        \State $L \leftarrow majority(C)$
        \If {$size(L) = 1$}
            \State $location \leftarrow L$
        \Else
            \State $location \leftarrow RandomizeElement(L)$
        \EndIf
        \State \textbf{return} $location$
\EndProcedure
\label{proposed_alg2}
\end{algorithmic}
\end{algorithm}

\subsection{ECU Authentication} \label{authentication}
The ECU authentication module is responsible for detecting unauthorized data transmissions on the CAN bus. For each legitimate ECU \emph{i}, a binary classifier is built based on a CNN. 

\textbf{CNN binary classifier architecture.} The following settings are used:
\begin{itemize}
    \item The classifier includes two convolutional layers followed by a max pooling layer to reduce the size. Each convolution layer has 32 filters.
    \item One fully connected layer is attached, which contains 100 neurons.
    \item All layers use the rectified linear unit (ReLU) as an activation function. 
    \item A sigmoid layer with a single unit is attached; this layer is aimed at producing the probability that a given example is associated with ECU \emph{i}.
\end{itemize}

\textbf{Training set.} The voltage signals transferred on the CAN bus and associated with the legitimate ECUs to authenticate. To train the binary classifier for ECU \emph{i}, each signal is classified according to the associated frame's origin ('1' if the origin of the signal is ECU \emph{i} and '0' otherwise).

\textbf{Training phase.} 
To address a possible data unbalance, we use the cost-sensitive learning method described in ~\cite{7727770}. The idea behind this method is that the training procedure is modified so that some examples have more or less errors than others.
In addition, to avoid overfitting, we define two dropouts set at 0.5; one is for the max pooling layer, and the other is for the fully connected layer.

During the training phase of each binary classifier, we use the RMSProp optimizer, with a learning rate of 0.0001., and \emph{binary cross-entropy} is used as the loss function. First, we chronologically extract 30\% of the training set to serve as the validation set. Then, we train the network until the loss function reaches its minimum on the validation set. 

\textbf{Authentication phase.} Given a signal associated with a CAN frame, we extract its ID and apply the appropriate binary classifier to the signal. The output returned from the classifier is the probability that the given signal matches the CAN frame ID. If the network output is less than 0.5, an alert is generated, and the ECU identification module is used to return the real origin of the CAN frame.

\subsection{ECU Identification}
The ECU Identification module focuses on identifying the real origin of a CAN frame. Its main building block is the binary classifiers which are generated as part of the Continuous ECU Authentication module. Each binary classifier is associated with one ECU. Thus, given a signal, we call each of the binary classifiers and return the appropriate identifier according to highest value returned. If the highest value is less than 0.5, and the CAN bus is compromised, we conclude that this signal is associated with a new device introduced on the CAN bus.

\section{\label{sec:results}Experiments and Results}

\subsection{Evaluation Setups}
\subsubsection{CAN Bus Prototype} 
\begin{figure}[th!]
\scriptsize
\centering
\includegraphics[width=0.47\textwidth]{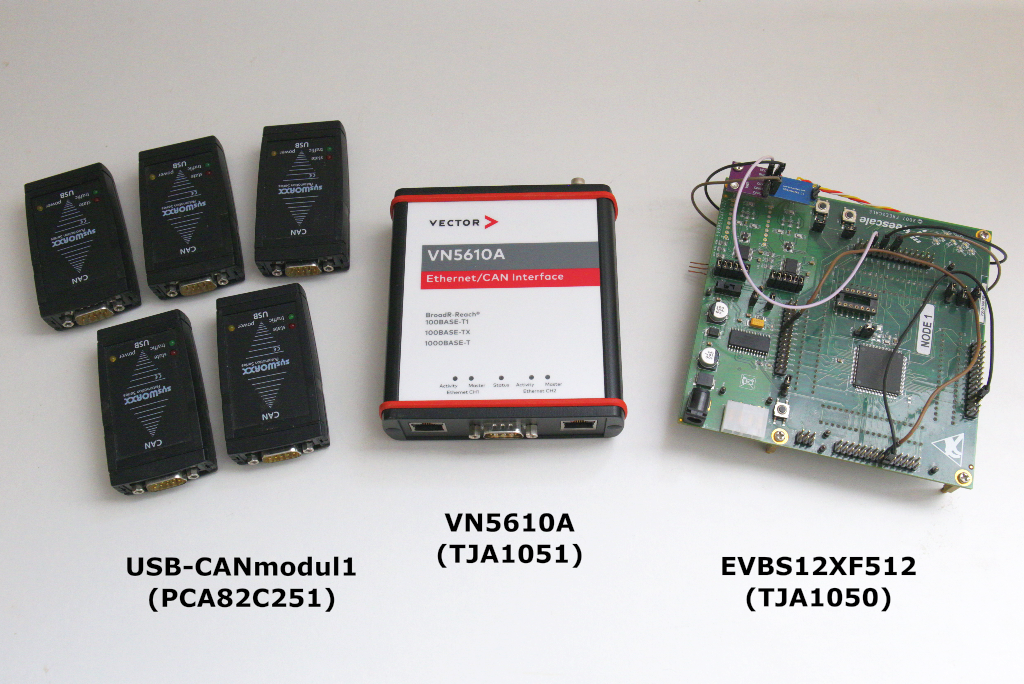}
\vspace{0.5cm}
\caption{CAN bus prototype.}
\label{fig:setup}
\end{figure}

As shown in Figure \ref{fig:setup}, our experimental setup is identical to the setup that was used in prior research ~\cite{murvay2020tidal}. In this section, we show that significantly better results are achieved when using the proposed physical intrusion detection and localization mechanism.

The CAN bus prototype is also used for the evaluation of the proposed ECU authentication and identification mechanism.

\begin{figure}[th!]
\scriptsize
\centering
\includegraphics[width=0.47\textwidth]{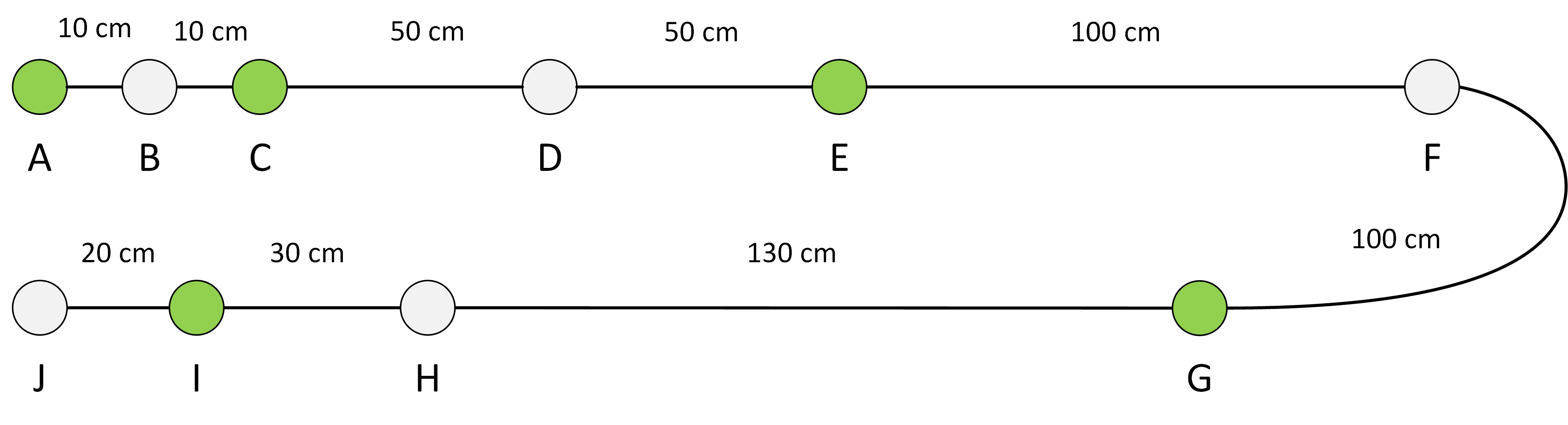}
\vspace{0.5cm}
\caption{Location of legitimate ECUs (green) and open entry points (gray) for intruders on our CAN bus prototype.}
\label{fig:experimental_bus}
\end{figure}

\textbf{Network configurations.} As illustrated in Figure \ref{fig:experimental_bus}, 10 connection points are located on the CAN bus. Some of them (green) are for legitimate ECUs, and the others (gray) are left open for malicious ECUs to be connected to the CAN bus.

\begin{figure}[th!]
\scriptsize
\centering
\includegraphics[width=0.47\textwidth]{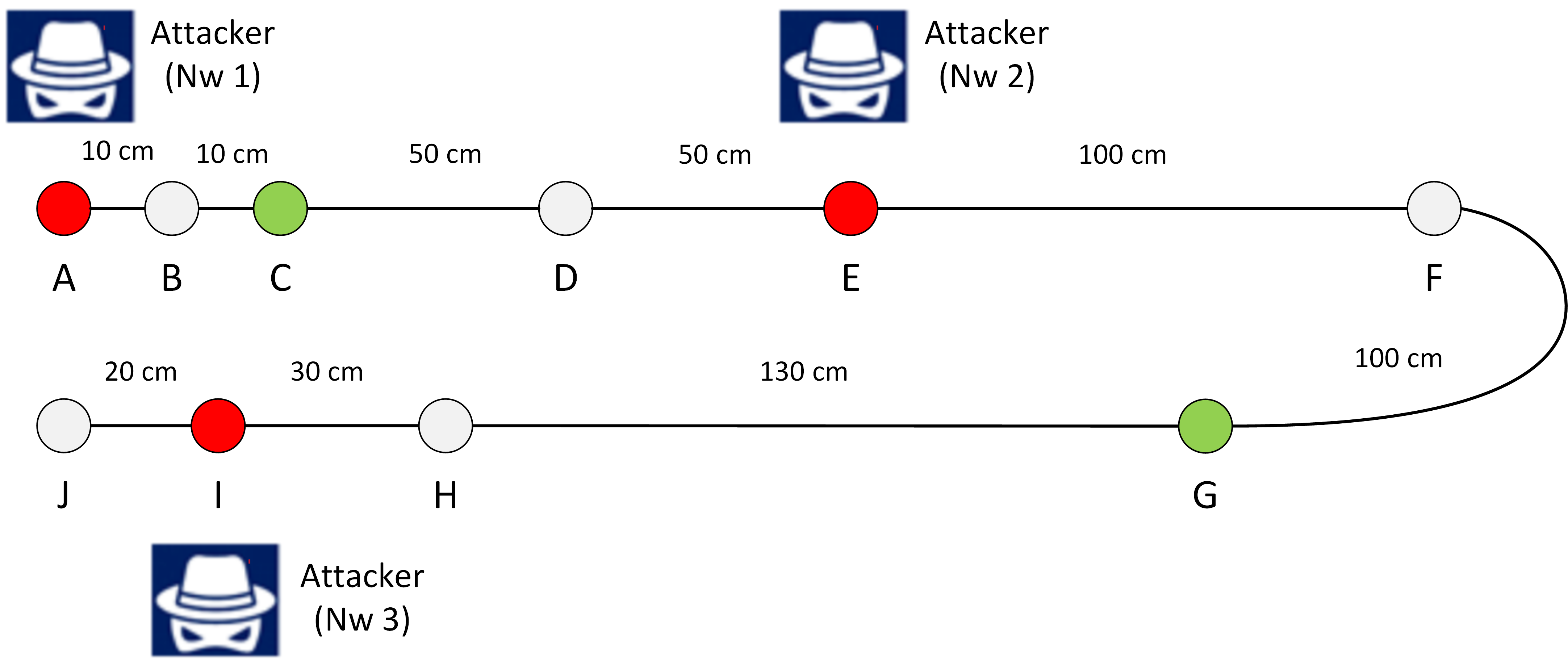}
(i) \emph{dirty} network configurations with replaced nodes (Nw 1-3).
\vspace{0.5cm}

\includegraphics[width=0.47\textwidth]{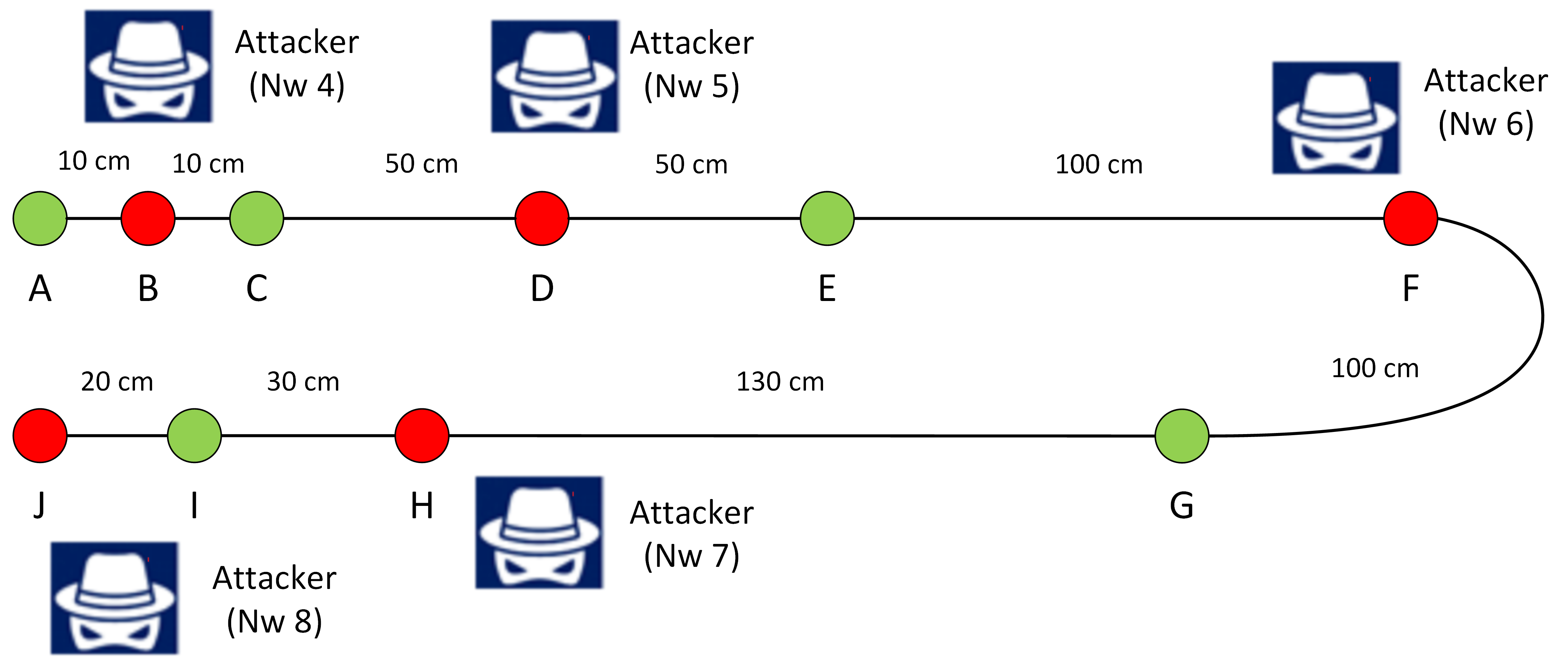}
(ii) \emph{dirty} network configurations with inserted nodes (Nw 4-8).
\vspace{0.5cm}

\caption{The adversarial network configurations examined.}
\label{fig:experimental_bus_adv}
\end{figure}

A number of network conﬁgurations are built using the available ECUs, in order to provide a range of speciﬁc test cases on which to evaluate the \MethodName system:
\begin{itemize}
    \item \emph{Network 0} - a \emph{clean} network in which all of the legitimate ECUs (and only those ECUs) are connected and transfer CAN frames, as depicted in Figure \ref{fig:experimental_bus}. 
    \item \emph{Networks 1-3} - \emph{dirty} networks in which a malicious ECU replaces a legitimate ECU at one of the locations depicted by the red circles in Figure \ref{fig:experimental_bus_adv} (i).
    \item \emph{Networks 4-8} - \emph{dirty} networks in which a malicious ECU is inserted into the CAN bus at one of the locations depicted by the red circles in Figure \ref{fig:experimental_bus_adv} (ii).
\end{itemize}

\textbf{Specific ECUs and their placement.} As depicted in Figure \ref{fig:setup}, we employ PC-to-CAN adapters (USB-CANmodul1 and VN5610A) and the EVBS12XF512 automotive grade development board, equipped with an external transceiver (TJA1050), to build our setup. For each ECU, Table \ref{tab:experiment_devices} lists the assigned abbreviated notation, the device type, the transceiver type, the amount, and the role in our experiment. $L_i$ is the legitimate ECU \emph{i} 
$(1 \leq i \leq 5)$, $A_1$ stands for the malicious ECU used for training, and $A_2$ (a completely different ECU related to $A_1$) stands for the malicious ECU used for testing. The network conﬁgurations, along with their designations, are listed in Table \ref{tab:experiment_confs}.

\begin{table}[th!]
\centering
\scriptsize
\caption{ECU devices and transceiver types and their role in the experiments.}
 \begin{tabular}{||c c c c c||} 
 \hline
 Abbrev. & Device & Transceiver & Amount & Role \\ [0.5ex] 
 \hline\hline
 ${L_i}$ & USB-CANmodul1  &  PCA82C251 & 5 & legitimate \\ 
 \hline
  ${A_1}$ & VN5610A & TJA1051 & 1 & adversary \\
 \hline
  ${A_2}$ & EVBS12XF512 & TJA1050 & 1 & adversary \\  [1ex] 
 \hline
\end{tabular}
\label{tab:experiment_devices}
\end{table}

\begin{table}[h!]
\centering
\scriptsize
\caption{Experimental network conﬁgurations.}
\resizebox{\columnwidth}{!}{\begin{tabular}{|c|c|l|c|l|c|l|c|l|c|l|}
\hline
\multirow{2}{*}{\begin{tabular}[c]{@{}c@{}}Nw.\\ Conf.\end{tabular}} & \multicolumn{10}{c|}{Connection point}                                                                                                                        \\ \cline{2-11} 
                                                                     & A    & \multicolumn{1}{c|}{B} & C    & \multicolumn{1}{c|}{D} & E    & \multicolumn{1}{c|}{F} & G    & \multicolumn{1}{c|}{H} & I    & \multicolumn{1}{c|}{J} \\ \hline
Nw0                                                                  & $L_{1}$ & \multicolumn{1}{c|}{}  & $L_{2}$ & \multicolumn{1}{c|}{}  & $L_{3}$ & \multicolumn{1}{c|}{}  & $L_{4}$ & \multicolumn{1}{c|}{}  & $L_{5}$ & \multicolumn{1}{c|}{}  \\ \hline
Nw1                                                                  & $A_{1,2}$ & \multicolumn{1}{c|}{}  & $L_{2}$ & \multicolumn{1}{c|}{}  & $L_{3}$ & \multicolumn{1}{c|}{}  & $L_{4}$ & \multicolumn{1}{c|}{}  & $L_{5}$ & \multicolumn{1}{c|}{}  \\ \hline
Nw2                                                                  & $L_{1}$ & \multicolumn{1}{c|}{}  & $L_{2}$ & \multicolumn{1}{c|}{}  & $A_{1,2}$ & \multicolumn{1}{c|}{}  & $L_{4}$ & \multicolumn{1}{c|}{}  & $L_{5}$ & \multicolumn{1}{c|}{}  \\ \hline
Nw3                                                                  & $L_{1}$ & \multicolumn{1}{c|}{}  & $L_{2}$ & \multicolumn{1}{c|}{}  & $L_{3}$ & \multicolumn{1}{c|}{}  & $L_{4}$ & \multicolumn{1}{c|}{}  & $A_{1,2}$ & \multicolumn{1}{c|}{}  \\ \hline
\multicolumn{1}{|l|}{Nw4}                                            & $L_{1}$ & $A_{1,2}$                   & $L_{2}$ &                        & $L_{3}$ &                        & $L_{4}$ &                        & $L_{5}$ &                        \\ \hline
\multicolumn{1}{|l|}{Nw5}                                            & $L_{1}$ &                        & $L_{2}$ & $A_{1,2}$                   & $L_{3}$ &                        & $L_{4}$ &                        & $L_{5}$ &                        \\ \hline
\multicolumn{1}{|l|}{Nw6}                                            & $L_{1}$ &                        & $L_{2}$ &                        & $L_{3}$ & $A_{1,2}$                   & $L_{4}$ &                        & $L_{5}$ &                        \\ \hline
\multicolumn{1}{|l|}{Nw7}                                            & $L_{1}$ &                        & $L_{2}$ &                        & $L_{3}$ &                        & $L_{4}$ & $A_{1,2}$                   & $L_{5}$ &                        \\ \hline
\multicolumn{1}{|l|}{Nw8}                                            & $L_{1}$ &                        & $L_{2}$ &                        & $L_{3}$ &                        & $L_{4}$ &                        & $L_{5}$ & $A_{1,2}$                   \\ \hline
\end{tabular}}
\label{tab:experiment_confs}
\end{table} 

\subsubsection{Real vehicle} 
\begin{figure}[th!]
\scriptsize
\centering
\includegraphics[width=0.47\textwidth]{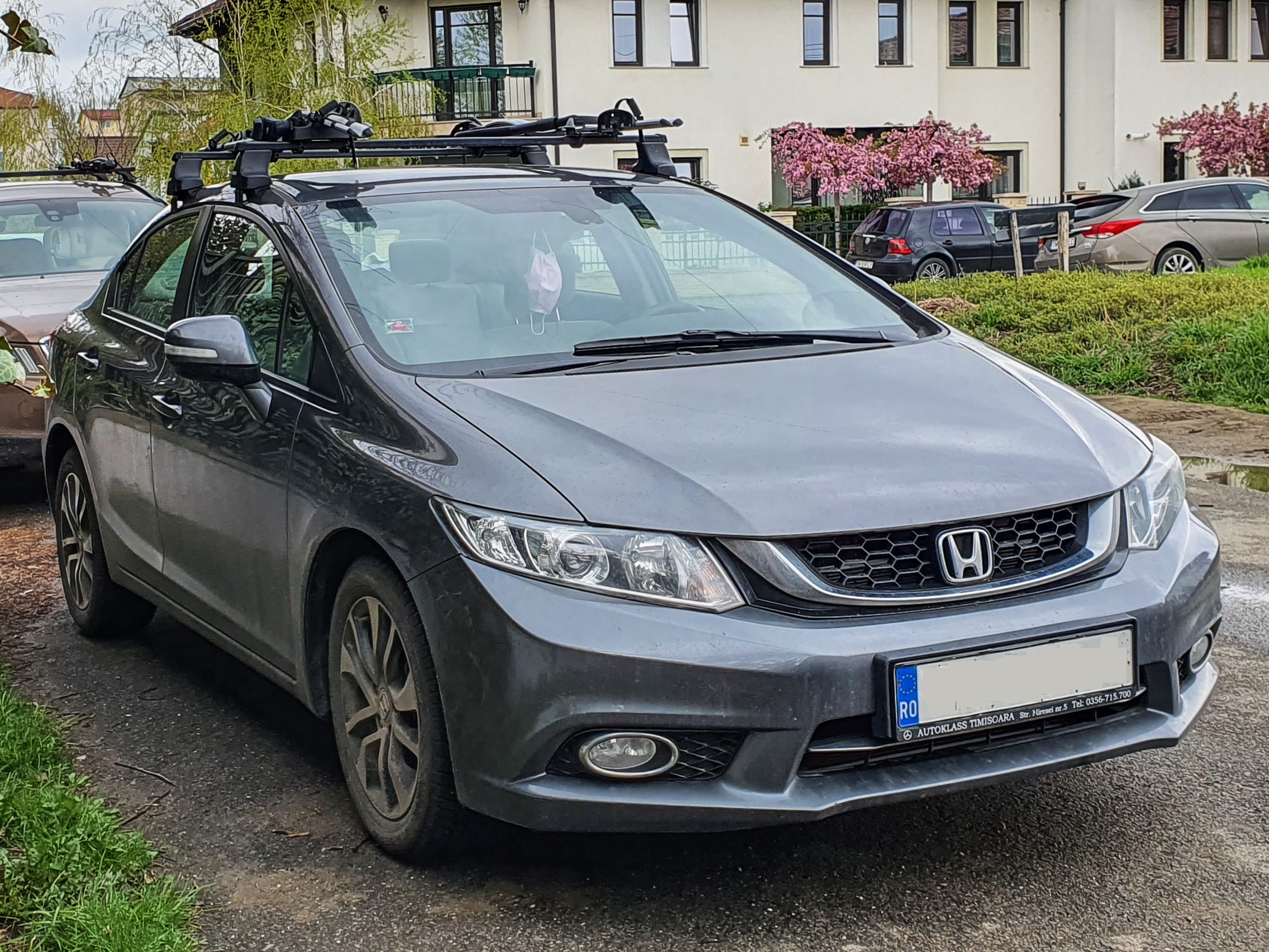}
\vspace{0.5cm}
\caption{2015 Honda Civic.}
\label{fig:setup_car}
\end{figure}

One car, a 2015 Honda Civic (Figure \ref{fig:setup_car}), was used to evaluate the robustness of the proposed ECU authentication method. Through the OBD-II port, the voltage signals were sampled from the in-vehicle CAN bus containing six ECUs, running at 500 Kbps.

\subsection{Results}

\subsubsection{Evaluation of the Physical Intrusion Detection Module}
As described in the previous section, this module focuses on detecting whether the CAN bus is compromised or not. The CAN bus prototype (Figure \ref{fig:setup}) is used to evaluate this module. For training and evaluation, we sample CAN-H values only.

\textbf{Training set collection.} Hundreds of signals are collected to train the autoencoder, all of which are collected from network 0. A detailed description of the training procedure was provided in the previous section.

\textbf{Test set collection.} Thousands of signals are collected to test the autoencoder. All of which are collected from network 0, and the expected prediction for each signal is \emph{clean}. Thousands more signals are collected from networks 1-8, and the expected prediction for each of those signals is \emph{dirty}. The malicious ECUs used for insertion and replacement are ${A_1}$ and ${A_2}$.

\begin{figure}[h]
\centering
\includegraphics[width=0.47\textwidth]{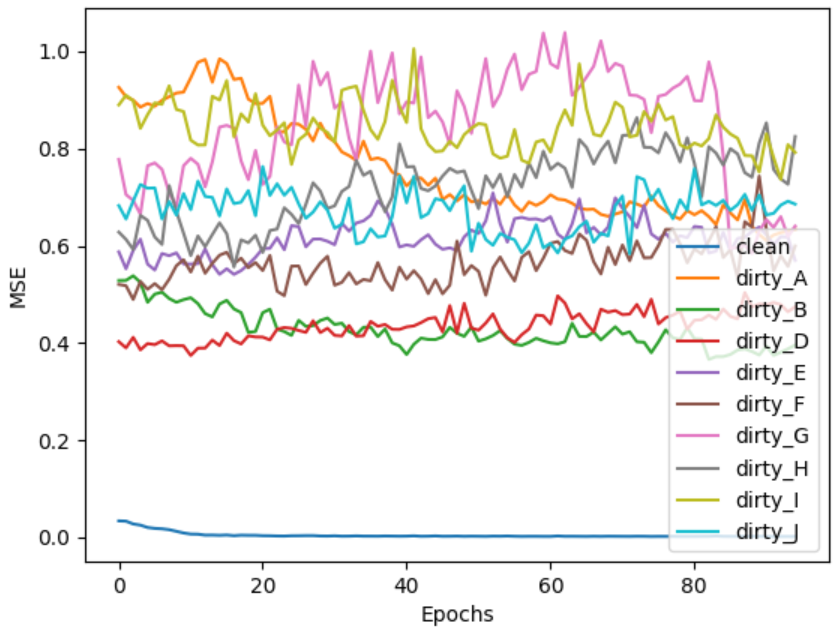}
\caption{The average MSE of \emph{clean} and \emph{dirty} signals as a function of the number of autoencoder training epochs.}
\label{fig:autoencoder_training}
\end{figure}

\textbf{Detection evaluation.} Figure \ref{fig:autoencoder_training} presents the average MSE of \emph{clean} and \emph{dirty} signals as a function of the number of autoencoder training epochs. As can be seen, there is a large margin between the reconstruction errors of \emph{clean} and \emph{dirty} scenarios. Unsurprisingly, our evaluation results show 100\% detection accuracy.
The sampling frequency used in this experiment was 125 MS/s.

\subsubsection{Evaluation of the Intrusion Point Localization Module}
As described in the previous section, this module is responsible for physically locating the intruder on the CAN bus when it has been compromised. The CAN bus prototype (Figure \ref{fig:setup}) was used to evaluate this module. For training and evaluation, we sample CAN-L values only.

Since replacement point localization relies on the ability to authenticate the legitimate ECUs, its performance is derived from the evaluation performed on the authentication method. The evaluation results of the authentication method for both the CAN bus prototype and a real vehicle are presented in the next section. We now focus on the evaluation results of the insertion point localization module. 

\textbf{Training set collection.} Thousands of signals are collected from networks 4-8 (hundreds per legitimate ECU) and assigned respectively with points B, D, F, H, and J (see Figure \ref{fig:experimental_bus_adv}). For each ECU, hundreds of signals per each point are collected.
Those signals are provided to the data augmentation algorithm (Algorithm 2, the $S^i$ parameter) which generates more signal examples for training. To generate the entire dataset for training, Algorithm 2 is executed five times, once against one legitimate ECU that the CAN bus prototype contains. 

On a call \emph{i} to Algorithm 2, we provide the collected signals associated with ECU \emph{i} as input (denoted by $S^i$). Another input to the algorithm is the set of the insertion points $P$={B, D, F, H, J}, a parameter $K$ set at 20, and a parameter $R$ set at 10. The resulting signals are used to train the VGG16 multiclass classifier (the training procedure was described in the previous section). The malicious ECU used for insertion and replacement is ${A_1}$.

\textbf{Test set collection.} Thousands of signals are collected from networks 4-8. For each ECU, hundreds of signals per each point are collected. The malicious ECU used for insertion and replacement is ${A_2}$. 

\textbf{Localization evaluation.}
The localization evaluation is performed by providing the multiclass classifier with five signals as input (one per legitimate ECU). Then, a majority is executed to return a final prediction, as described in the previous section (Algorithm 3). 
As can be seen in Table \ref{tab:node_location_classification_maj}, excellent results were achieved.
These results reflect the ability of the proposed module to localize inserted intruder using legitimate ECUs' signals only.
The sampling frequency used in this experiment was 500 MS/s.

\begin{table}[thb!]
\centering
\small
\setlength\tabcolsep{1.7pt}
\caption{Confusion matrix of the proposed insertion localization module.}
\begin{center}
\begin{tabular}{@{}cc||*{5}{c|}}
\multicolumn{1}{c}{}  &   &\multicolumn{5}{|c|}{Predicted}\\
\multicolumn{1}{c}{} & &B    &D    &F    &H    &J   \\ \hline \hline
\multirow{5}*{\rotatebox{90}{Actual}}  
   & B &100  &0    &0    &0    &0      \\ \cline{2-7} 
   & D &0  &100    &0    &0    &0      \\ \cline{2-7}
   & F &0  &0    &100    &0    &0      \\ \cline{2-7}
   & H &0  &0    &0    &100    &0      \\ \cline{2-7}
   & J &0  &0    &0    &0    &100     \\ \cline{2-7}
\end{tabular}
\end{center}
\label{tab:node_location_classification_maj}
\end{table}

\begin{table}[t!]
\centering
\scriptsize
\setlength\tabcolsep{2pt}
\caption{Authentication experiment results evaluated on a CAN bus prototype. \label{tab:authentication_prototype}}
\begin{center}
\begin{tabular}{|c|c|c|c|c|c|c|c|c|c|}
\hline
\multicolumn{2}{|c|}{ECU1} & \multicolumn{2}{c|}{ECU2} & \multicolumn{2}{c|}{ECU3} & \multicolumn{2}{c|}{ECU4} & \multicolumn{2}{c|}{ECU5} \\ \hline
FRR          & FAR         & FRR         & FAR         & FRR          & FAR        & FRR          & FAR        & FRR          & FAR        \\ \hline
\multicolumn{10}{|c|}{\textbf{clean network}}                                    \\ \hline
0        & 0       & 0       & 0           & 0        & 0          & 0        & 0          & 0       & 0          \\ \hline
\multicolumn{10}{|c|}{\textbf{dirty ($A_1$ is silent)}}  

\\ \hline
0.002        & 0       & 0.001       & 0           & 0        & 0          & 0        & 0          & 0       & 0          \\ \hline
\multicolumn{10}{|c|}{\textbf{dirty ($A_2$ is silent)}}                  

              \\ \hline
0.08         & 0       & 0           & 0       & 0        & 0          & 0        & 0          & 0        & 0          \\ \hline
\multicolumn{10}{|c|}{\textbf{dirty ($A_1$ is active)}} 

\\ \hline
0.002         & 0       & 0          & 0.001       & 0        & 0          & 0        & 0          & 0        & 0          \\ \hline
\multicolumn{10}{|c|}{\textbf{dirty ($A_2$ is active)}}                                                                                                               \\ \hline
0.08         & 0        & 0        & 0       & 0       & 0       & 0       & 0          & 0        & 0.04          \\ \hline
\end{tabular}
\end{center}
\end{table}

\begin{table}[t!]
\centering
\scriptsize
\setlength\tabcolsep{1.1pt}
\caption{Authentication experiment results evaluated on a real vehicle. \label{tab:authentication_honda}}
\begin{center}
\begin{tabular}{|c|c|c|c|c|c|c|c|c|c|c|c|}
\hline
\multicolumn{2}{|c|}{ECU1} & \multicolumn{2}{c|}{ECU2} & \multicolumn{2}{c|}{ECU3} & \multicolumn{2}{c|}{ECU4} & \multicolumn{2}{c|}{ECU5} & \multicolumn{2}{c|}{ECU6}\\ \hline
~FRR~           & ~FAR~        & ~FRR~         & ~FAR~         & ~FRR~         & ~FAR~    & ~FRR~ & ~FAR~ & ~FRR~ & ~FAR~ & ~FRR~ & ~FAR~   \\ \hline
\multicolumn{12}{|c|}{\textbf{0 minutes}}                                                    \\ \hline
 0             & 0          & 0           & 0           & 0.002       & 0      & 0             & 0          & 0       & 0           & 0.003       & 0      \\ \hline
\multicolumn{12}{|c|}{\textbf{15 minutes}}                                                   \\ \hline
0.008         & 0           & 0            & 0.003          & 0.006       & 0.003       & 0          & 0           & 0         & 0        & 0.005       & 0    \\ \hline
\multicolumn{12}{|c|}{\textbf{30 minutes}}                                                   \\ \hline
0          & 0           & 0            & 0        & 0        & 0    & 0          & 0           & 0         & 0            & 0.005        & 0            \\ \hline
\multicolumn{12}{|c|}{\textbf{60 minutes}}                                                   \\ \hline
0           & 0           & 0         & 0.002           & 0        & 0.003     & 0          & 0           & 0         & 0            & 0.008        & 0       \\ \hline
\end{tabular}
\end{center}
\end{table}

\subsubsection{Evaluation of the ECU Authentication Module}
As described in the previous section, this module focuses on detecting unauthorized data transmissions on the CAN bus. Both the CAN bus prototype and a real vehicle were used to evaluate this module. We found that using the differential between the CAN-H and CAN-L values contributes to the robustness of the proposed ECU authentication module.

\textbf{Training set collection (CAN bus prototype).} 
To train the binary classifiers, hundreds of signals (legitimate only) are collected from networks 0-8. The malicious ECU used for insertion and replacement is ${A_1}$. 
 
\textbf{Test set collection (CAN bus prototype).} To evaluate the binary classifiers, thousands of signals (legitimate and non-legitimate) are collected from networks 0-8. 
The malicious ECUs used for insertion and replacement are ${A_1}$ and ${A_2}$.

\textbf{Classification evaluation (CAN bus prototype).} Each binary classifier's performance is evaluated in terms of the false rejection rate (FRR) and false acceptance rate (FAR). As illustrated in Table \ref{tab:authentication_prototype}, we evaluate each classifier's performance separately in three scenarios. First, we evaluate each classifier's performance on a \emph{clean} CAN bus only (network 0). Then, we evaluate each classifier's performance in a \emph{dirty} scenario in which the malicious ECU is silent (networks 1-8, excluding the malicious ECU's signals). Finally, we evaluate each classifier's performance in a \emph{dirty} scenario in which the malicious ECU is active (networks 1-8, including the malicious ECU's signals).

As can be seen in Table \ref{tab:authentication_prototype}, excellent results were achieved for both the \emph{clean} network and the \emph{dirty} network with silent intruders. In addition, good results were achieved when the signals generated by $A_1$ and $A_2$ were used. 
The sampling frequency used in this experiment was 500 MS/s.

\textbf{Training set collection (real vehicle).} To train the binary classifiers, several thousands of signals are collected when the vehicle is turned on. All of them are collected while the vehicle is stationary.

\textbf{Test set collection (real vehicle).} To evaluate the robustness of the binary classifiers, signals are collected while the vehicle is \textbf{moving}. 
The signals are grouped into four separate datasets according to the length of time the car has been running: (i) 0 minutes (immediately after the car was started), (ii) 15 minutes, (iii) 30 minutes, and (iv) 60 minutes.
Each group contains thousands of signals for each ECU.

\textbf{Classification evaluation (real vehicle).} We evaluate each classifier's performance in terms of the FRR and FAR, separately on each dataset.
Table \ref{tab:authentication_honda} presents the performance of the proposed method on each dataset. We can see that a low FRR and FAR were achieved on each dataset. 
Since the fingerprints are generated based on signals collected for just a few minutes once the car has started, we conclude that the results achieved indicate the robustness of our method to vibrations and temperature variations.
The sampling frequency used in this experiment was 250 MS/s.

\section{\label{sec:deployment}System Deployment}

Similar to the mechanism proposed in~\cite{canary2021bogdan}, our system can be implemented on an external node attached to the CAN bus. 
This proposed deployment can address the large number of vehicles that are already on the road.
The dataset required to induce the models of the system for vehicles on the road can be collected at the garage.
For new vehicles, the dataset required to induce the system's models can be collected after the vehicle has been produced, i.e., during the vehicle testing phase on the production line. To achieve the accurate detection demonstrated in Section~\ref{sec:results}, a DSP with a sampling rate of 500 MS/s should be used in the deployed system. 

As for the computational power, 
on a 2.11GHz Intel Core i7-8665U processor, it took about three seconds to parse 15K frames during authentication, which corresponds to a processing time in the order of $200 \mu s$ per frame. This corresponds to the time required to process frames in real time since the time spent by a frame on a 500 Kbps CAN bus is around $200 \mu s$. This amount of computational power is available on a modern high-end DSP.
The identification evaluation performed on the one-hour experiment on the Honda Civic did not require retraining when a sufficient number of samples (i.e., around 1,000 per ECU) was provided when the vehicle was started. This number of samples can usually be collected from an ECU in a matter of seconds or minutes. 
Moreover, since all of the models presented in this work are based on neural networks, which are known to be adaptive and thus support online training, only a small amount of data can be stored in the memory at any given time.

As stated in Section~\ref{sec:intro}, our proposed system complements a prevention mechanism proposed in ~\cite{canary2021bogdan} that requires accurate localization of the intrusion point, which our proposed system facilitates by using deep learning. From a data collection perspective, the mechanism described by the authors in ~\cite{canary2021bogdan} can also be used for the automatic examination and diagnosis of specific segments of the CAN bus.

\section{\label{sec:conclusion}Summary and Conclusion}
In this study, we demonstrate how CAN bus voltage signals can be used to identify and locate unauthorized topology changes on the CAN bus network with high accuracy. Since we do not depend on an adversary's transmission, our physical intrusion detection and localization mechanism is effective against silent intruders.

Methods proposed in other studies that used an identical setup but were based on timing analysis were unable to localize the intruder in cases in which a new ECU was inserted into the CAN bus. By using deep learning with data augmentation, we are able to localize the intruder, even when a new (unknown) ECU is connected to the CAN bus.

In addition, we show that the proposed mechanism can successfully authenticate ECUs, even when the network topology has changed.
We demonstrate that our proposed authentication module allows us to identify the legitimate ECUs on a variety of network topologies (e.g., when a new ECU is introduced on the bus or replaces an existing one). 

Using a real vehicle, we also show that our proposed authentication mechanism is robust to environmental changes. We demonstrate this under the following demanding conditions: training using data collected while the vehicle is stationary and testing over a long period of time (over an hour) when the vehicle is moving. 
Since we rely on electrical properties, which are unique to each ECU, spoofing attacks are largely infeasible.

The high identification accuracy obtained in the real vehicle evaluation indicates that the neural network created for the authentication task can generalize and does not overfit certain environmental conditions. 
Moreover, the fact that an identical authentication method works well in two different demanding environments (a real vehicle and a CAN bus prototype), together with neural networks' adaptive property, indicate that our proposed system can be easily transformed into a plug and play solution. No hyperparameter tuning is required.

In future research, we plan to test the proposed system in additional scenarios, e.g., when one ECU goes into bus off or low-power mode, when the supply voltage from the ECUs' fluctuates, and when other distinct types of ECUs are added to the CAN bus by the attacker.

\bibliographystyle{IEEEtranS}
\bibliography{main}

\clearpage 


\end{document}